\documentstyle[12pt,titlepage]{article}  
\newcommand{\nin}{\noindent}
\newcommand{\be}{\begin{equation}}
\newcommand{\ee}{\end{equation}}
\newcommand{\bea}{\begin{eqnarray}}
\newcommand{\eea}{\end{eqnarray}}
\newcommand{\br}{\hskip .25cm/\hskip -.25cm}

\newcommand{\nonu}{\nonumber\\}
\newcommand{\la}{\hskip .25cm^\leftarrow\hskip -.25cm}

\newcommand{\ol}{\overline}

\begin{document}

\hfill {NTUA-1/01}

\vspace{2cm}

\begin{center}
{\Large QED in external fields,\\
a functional point of view}

\vspace{1cm}

J.Alexandre{\footnote {jalex@central.ntua.gr}}

Department of Physics, National Technical University of
Athens,

Zografou Campus, 157 80 Athens, Greece

\vspace{2cm}

Abstract
\end{center}

A functional partial differential equation is set for the proper
graphs generating functional of QED in external electromagnetic
fields. This equation leads to the evolution of the proper graphs
with the external field amplitude and the external field gauge
dependence of the complete fermion propagator and vertex is
derived non-perturbatively.

\vspace{2cm}

The study of QED in the presence of external electromagnetic
fields started long ago with the computation of the quantum
corrections to the Maxwell Lagrangian \cite{euler}, for which
recent and rich studies can be found in \cite{schubert}. We find
also among the studies of the effect of an external field the
dynamical chiral symmetry breaking by a magnetic field, the
magnetic catalysis \cite{miransky}.

In this framework,
we will describe here a non-perturbative approach, similar to
the background field methods \cite{abbott}, which will lead us
to the external field gauge dependence of the full fermion
propagator and the full vertex. For this we will set up
a functional differential equation showing
the evolution of the effective action (Legendre
transform of the connected graphs generator
functional)
with the amplitude of the external field. The interesting point is that
this differential equation is exact and thus independent of pertubative
expansions.

The method described here
is similar to the one given in \cite{jj} where the differential equation
described the evolution of the effective action
with the mass scale of a scalar theory and lead to the well-known
one-loop effective action after integration.

We will discuss here the dependence of the proper functions on the
gauge of the external electromagnetic field and not of the
dynamical one. The gauge dependence with respect to the dynamical
field is discussed in \cite{landau}. The authors take a usual
covariant gauge fixing term and show the dependence of the proper
functions on the gauge parameter. Here we will not write this
gauge fixing term, although necessary to define the path integral,
so as to focus on the effects of the external field. The
Lagrangian is

\bea
{\cal L}=-\frac{1}{4}{\cal F}_{\mu\nu}{\cal F}^{\mu\nu}+
\ol\Psi \left(i\br \partial-e\br{\cal A}-g\br A^{ext}-m\right)\Psi
\eea

\nin where ${\cal A}_\mu$ is the dynamical gauge field of strength
${\cal F}_{\mu\nu}$ and $A_\mu^{ext}$ the classical (external)
electromagnetic field. $e$ is the QED coupling and $g$ the
coupling to the external field which is taken different form $e$
so that it can control the amplitude of $A_\mu^{ext}$.

We will introduce the effective action $\Gamma$ of the model,
defined as the Legendre transform of the connected graphs
generator functional
and our aim will be to derive the exact functional equation

\be\label{evolinit}
\partial_g\Gamma={\cal G}[\Gamma]
\ee

\nin from which we can extract the evolution
of the proper graphs.

In what follows, the Lorentz, Dirac and space-time indices will
not be explicitly written if not necessary. We will note 'tr' the
trace over Dirac indices and 'Tr' the trace over Dirac and
space-time indices. The computations will be done in dimension
$d$.

The connected graphs generator functional $W_g$ is given by

\bea\label{cggen}
\exp W_g[\ol\eta,\eta,j]
&=&Z_g[\ol\eta,\eta,j]\\
&=&\int{\cal D}[{\cal A},\ol\Psi,\Psi]
\exp\left\{i\int_x{\cal L}+i\int_x(j{\cal A}+\ol\eta\Psi+
\ol\Psi\eta)\right\}\nonumber
\eea

\nin where $\int_x=\int d^dx$. We note that we do not couple
the external field to the source $j$.
$W_g$ has the following functional derivatives

\bea\label{derivw}
\frac{\delta W}{\delta j}&=&\frac{1}{Z}\left<i{\cal A}\right>=iA\nonu
\frac{\delta W}{\delta\ol\eta}&=&\frac{1}{Z}\left<i\Psi\right>=i\psi\nonu
W\frac{\la\delta}{\delta\eta}&=&\frac{1}{Z}\left<i\ol\Psi\right>=i\ol\psi\nonu
\frac{\delta}{\delta\ol\eta}W\frac{\la\delta}{\delta\eta}&=&-\ol\psi\psi
+\frac{1}{Z}\left<\ol\Psi\Psi\right>
\eea

\nin where the expectation value $\left<{\cal O}\right>$ of an operator
${\cal O}$ is

\be
\left<{\cal O}\right>=\int{\cal D}[{\cal A},\ol\Psi,\Psi]
~{\cal O}~\exp\left\{i\int_x{\cal L}+
i\int_x(j{\cal A}+\ol\eta\Psi+\ol\Psi\eta)\right\}
\ee

\nin Inverting the relations between $(j,\ol\eta,\eta)$ and $(A,\ol\psi,\psi)$,
we define the effective action $\Gamma_g[A,\ol\psi,\psi]$ as the
Legendre transform
of $W_g[j,\ol\eta,\eta]$ by

\be
W_g=i\Gamma_g+i\int_x\left(jA+\ol\eta\psi+\ol\psi\eta\right)
\ee

\nin From this definition we extract the following functional derivatives:

\bea\label{derivg}
\frac{\delta\Gamma}{\delta A}&=&-j\nonu
\frac{\delta\Gamma}{\delta\ol\psi}&=&-\eta\nonu
\Gamma\frac{\la\delta}{\delta\psi}&=&-\ol\eta\nonu
\frac{\delta}{\delta\ol\psi}\Gamma\frac{\la\delta}{\delta\psi}&=&
-\frac{\delta\ol\eta}{\delta\ol\psi}
=-i\left(\frac{\delta}{\delta\ol\eta}W\frac{\la\delta}{\delta\eta}\right)^{-1}
\eea

\nin The evolution equation with $g$ of the connected graphs
generator functional is, according to (\ref{derivw})

\bea\label{derivtw}
\partial_gW&=&\frac{1}{Z}\left<-i\int_x\ol\Psi\br A^{ext}\Psi\right>\nonu
&=&-i\int_x\left(\ol\psi\br A^{ext}\psi+\br A^{ext}
\frac{\delta}{\delta\ol\eta}W\frac{\la\delta}{\delta\eta}
\right)
\eea

\nin To compute the evolution of $\Gamma_g$, one has to keep in mind
that the independent variables for this functional are $\ol\psi,\psi,A$
and $g$. Taking (\ref{derivw}) into account, we obtain then

\bea\label{evol1}
\partial_g\Gamma&=&-i\partial_gW-i\int_x\left(\partial_g\ol\eta
\frac{\delta W}{\delta\ol\eta}+
W\frac{\la\delta}{\delta\eta}\partial_g\eta+
\frac{\delta W}{\delta j}
\partial_g j\right)
-\int_x\left(\partial_gjA+\partial_g\ol\eta\psi+
\ol\psi\partial_g\eta\right)\nonu
&=&-i\partial_gW
\eea

\nin The relation (\ref{derivg}) between the second derivatives of
$W_g$ and $\Gamma_g$ implies then the exact functional evolution
equation for the effective action:

\be\label{evol}
\partial_g\Gamma+\int_x\ol\psi\br A^{ext}\psi=i\mbox{Tr}\left\{\br A^{ext}
\left(\frac{\delta}{\delta\ol\psi}\Gamma\frac{\la\delta}{\delta\psi}
\right)^{-1}\right\}
\ee

\nin We can actually write another evolution equation, which will
be more useful, using the equation of motion for the dynamical
gauge field $A_\mu$ that we now derive again as is done in
\cite{zuber}, so as not to miss any contribution.\\ We can assume
that the integral of a derivative vanishes, so that

\be
\int{\cal D}[{\cal A},\ol\Psi,\Psi]\frac{\delta}{\delta{\cal A}_\mu(x)}
\exp\left\{i\int{\cal L}+i\int(j{\cal A}+\ol\eta\Psi+\ol\Psi\eta)\right\}
=0
\ee

\nin which can be written

\be
\left(\Box g^{\mu\nu}-\partial^\mu\partial^\nu\right)
\left<{\cal A}_\nu\right>
-e\left<\ol\Psi\gamma^\mu\Psi\right>+Zj^\mu=0
\ee

\nin Using the relations (\ref{derivw}) and (\ref{derivg}), we find then

\be\label{wellknown}
\frac{\delta\Gamma}{\delta A_\mu}=
\left(\Box g^{\mu\nu}-\partial^\mu\partial^\nu\right)A_\nu
-e\ol\psi\gamma^\mu\psi+
ie~\mbox{tr}\left\{\gamma^\mu\left[\frac{\delta}{\delta\ol\psi}\Gamma
\frac{\la\delta}{\delta\psi}\right]^{-1}(x,x)\right\}
\ee

\nin Making the scalar product with $A^{ext}$, we obtain

\be\label{mvt}
\int_x A^{ext}_\mu\frac{\delta\Gamma}{\delta A_\mu}+
\frac{1}{2}\int_xF_{\mu\nu}^{ext}F^{\mu\nu}
+e\int_x\ol\psi \br A^{ext}\psi=
ie~\mbox{Tr}\left\{\br A^{ext}\left(\frac{\delta}{\delta\ol\psi}
\Gamma\frac{\la\delta}{\delta\psi}\right)^{-1}\right\}
\ee

\nin up to a surface term. (\ref{mvt}) and (\ref{evol})
give then a {\it linear} evolution
equation for the effective action $\Gamma$:

\be\label{evoll}
e\partial_g\Gamma=
\int_x A^{ext}_\mu
\frac{\delta\Gamma}{\delta A_\mu}+
\frac{1}{2}\int_xF_{\mu\nu}^{ext}F^{\mu\nu}
\ee

Let us check that $\partial_g\Gamma$ is invariant with respect to
an external gauge transformation $A^{ext}_\mu \to
A^{ext}_\mu+\partial_\mu\phi$. The second integral of the right
hand side of (\ref{evoll}) is of course invariant and the first
one becomes after an integration by parts where we disregard the
surface term

\be
\int_x A^{ext}_\mu\frac{\delta\Gamma}{\delta A_\mu}\to
\int_x A^{ext}_\mu\frac{\delta\Gamma}{\delta A_\mu}-
\int_x\phi\partial_\mu\left(\frac{\delta\Gamma}{\delta A_\mu}\right)
\ee

\nin But according to (\ref{derivg}), the additional integral
vanishes since

\be
\partial_\mu\left(\frac{\delta\Gamma}{\delta A_\mu}\right)=
-\partial_\mu j^\mu=0,
\ee

\nin due to the charge conservation.
Thus (\ref{evoll}) is gauge invariant, as expected.

To conclude with the effective action, we can give a general
solution of (\ref{evoll}) in terms of the proper graphs generator
functional without external field.

If we take the second derivative of $\Gamma$ with respect to $g$,
we obtain

\be
e^2\partial^2_g\Gamma=\int_{xy} A^{ext}_\mu(x)A^{ext}_\nu(y)
\frac{\delta^2\Gamma}{\delta A_\mu(x)\delta A_\nu(y)}
+\frac{1}{2}\int_xF_{\mu\nu}^{ext}F^{\mu\nu}_{ext}
\ee

\nin where we again disregarded the surface term.
We obtain in general, for $n\ge3$

\be
e^n\partial^n_g\Gamma=\int_{x_1...x_n}
A^{ext}_{\mu_1}(x_1)...A^{ext}_{\mu_n}(x_n)
\frac{\delta^n\Gamma}{\delta A_{\mu_1}(x_1)...\delta A_{\mu_n}(x_n)}
\ee

\nin Then we can make the resumation

\be
\Gamma_g=\sum_{n=0}^\infty\frac{g^n}{n!}\partial^n_g\Gamma_0
\ee

\nin and take $g=e$, which leads to the following
relation between $\Gamma$ (with the external field) and
$\Gamma_0$ (without the external field):

\be\label{rel}
\Gamma=\exp\left\{\int_x A^{ext}_\mu(x)\frac{\delta}{\delta A_\mu(x)}
\right\}\Gamma_0+\frac{1}{2}\int_xF_{\mu\nu}^{ext}F^{\mu\nu}
+\frac{1}{4}\int_xF_{\mu\nu}^{ext}F^{\mu\nu}_{ext}
\ee

\nin The integrals involving $F_{\mu\nu}^{ext}$ in (\ref{rel})
correspond to the subtraction of the kinetic contribution of the
external field which does not enter into account in the problem.

We recognize in (\ref{rel}) the functional translation operator
which is the generalization of

\be
\exp\left(x_0\frac{d}{dx}\right) f(x)=f(x+x_0)
\ee

\nin such that we can finally write

\be
\Gamma[\ol\psi,\psi,A]=\Gamma_0[\ol\psi,\psi,A+A^{ext}]
+\frac{1}{2}\int_xF_{\mu\nu}^{ext}F^{\mu\nu}
+\frac{1}{4}\int_xF_{\mu\nu}^{ext}F^{\mu\nu}_{ext}
\ee

\nin Thus the effective action of the theory with an external
field is the same functional as the one of the theory without
external field (but for the bare kinetic term), translated by the
vector $A^{ext}$ in the space of the functional variables. This
equivalence between the effective action with and without external
field is known in the background field methods \cite{abbott}.

Let us now come back to the equation (\ref{evoll}). Its
differentiation with respect to the fields
leads to the evolution of
the proper graphs. Let us take the second derivative
with respect to $\ol\psi$ and $\psi$ for vanishing sources.
We encounter then
the vertex function

\be\label{defvertex}
\Lambda^\mu(z;x,y)=-\frac{1}{e}\frac{\delta}{\delta A_\mu(z)}
\frac{\delta}{\delta\ol\psi(x)}
\Gamma\frac{\la\delta}{\delta\psi(y)}|_{A=\ol\psi=\psi=0}
\ee

\nin and the inverse fermion propagator

\be\label{definvprop}
G^{-1}(x,y)=-i\frac{\delta}{\delta\ol\psi(x)}\Gamma
\frac{\la\delta}{\delta\psi(y)}|_{A=\ol\psi=\psi=0}
\ee

\nin to obtain the following evolution equation:

\be\label{evolinvg}
\partial_g G^{-1}(x,y)=
i\int_zA^{ext}_\mu(z)\Lambda^\mu(z;x,y)
\ee

\nin To fix the idea, we can see that the tree-level proper graphs
$G^{-1}_{tree}$ and $\Lambda^\mu_{tree}$ verify (\ref{evolinvg})
since they are given by

\bea\label{bare}
G^{-1}_{tree}(x,y)&=&-i\left[i\br\partial_x-g\br A^{ext}(x)
-m\right]\delta(x-y)\nonu
\Lambda^\mu_{tree}(z;x,y)&=&\gamma^\mu\delta(x-z)\delta(y-z)
\eea

\nin The evolution equation
for the propagator is then obtained by noticing that

\be
\partial_g G^{-1}(x,y)=-\int_{z_1z_2}G^{-1}(x,z_1)
\partial_g G(z_1,z_2) G^{-1}(z_2,y)
\ee

\nin and therefore, according to (\ref{evolinvg})

\be\label{evolg}
\partial_gG(x,y)+i\int_{z_1z_2z_3}A^{ext}_\mu(z_1)
G(x,z_2)\Lambda^\mu(z_1;z_2,z_3)G(z_3,y)=0
\ee

Let us now perform a gauge transformation on the external field
$A^{ext}_\mu=A^0_\mu$ and write $A^1_\mu=A^0_\mu+\partial_\mu\phi$
(we will use the same notations for $G$ and $\Lambda^\mu$ corresponding
to the two gauges).
We obtain after an integration by parts where we omit the surface term

\bea
&&\partial_g G_1(x,y)+i\int_{z_1z_2z_3}A^0_\mu(z_1)
G_1(x,z_2)\Lambda^\mu_1(z_1;z_2,z_3)G_1(z_3,y)\nonu
&&=i\int_{z_1z_2z_3}\phi(z_1)
G_1(x,z_2)\partial_\mu^{z_1}\Lambda^\mu_1(z_1;z_2,z_3)G_1(z_3,y)
\eea

\nin Then the use of the Ward identity

\be\label{ward}
\partial_\mu^{z_1}\Lambda^\mu(z_1;z_2,z_3)=
\delta(z_2-z_1)G^{-1}(z_1,z_3)-\delta(z_3-z_1)G^{-1}(z_2,z_1)
\ee

\nin leads us to

\be\label{intermediaire}
\partial_g G_1(x,y)-i\Theta_{xy}G_1(x,y)
+i\int_{z_1z_2z_3}A^0_\mu(z_1)
G_1(x,z_2)\Lambda^\mu_1(z_1;z_2,z_3)G_1(z_3,y)=0
\ee

\nin where

\be
\Theta_{xy}=\int_x^y dz^\mu
\left[A^1_\mu(z)-A^0_\mu(z)\right]=\phi(y)-\phi(x)
\ee

\nin The equation (\ref{intermediaire}) can also be written

\bea
0&=&\partial_g\left\{G_1(x,y)e^{-ig\Theta_{xy}}\right\}\\
&&+i\int_{z_1z_2z_3}A^0_\mu(z_1)
\left\{G_1(x,z_2)e^{-ig\Theta_{xz_2}}\right\}
\left\{\Lambda^\mu_1(z_1;z_2,z_3)e^{-ig\Theta_{z_2z_3}}\right\}
\left\{G_1(z_3,y)e^{-ig\Theta_{z_3y}}\right\}\nonumber
\eea

\nin We recognize here the equation (\ref{evolg}) satisfied by
$G$ and $\Lambda^\mu$ in the gauge $A^0_\mu$ with the transformation law

\bea\label{transfo}
G_1(x,y)&=&G_0(x,y)e^{ig\Theta_{xy}}\nonu
\Lambda^\mu_1(z;x,y)&=&\Lambda^\mu_0(z;x,y)e^{ig\Theta_{xy}}
\eea

\nin which gives the gauge dependence of the proper graphs $G$ and
$\Lambda^\mu$.

Let us now turn to the constant field case where
we know from the
work by Schwinger \cite{euler} that the bare fermion propagator is
of the form

\be\label{phase}
G_{tree}(x,y)=e^{ig\int_x^y dz^\mu A^{ext}_\mu(z)}\tilde G_{tree}(x-y)
\ee

\nin where $\tilde G_{tree}$ depends on the difference $x-y$ only and
is gauge invariant since the gauge dependence of
$G_{tree}$ is contained in the phase (\ref{phase}), as can be seen from
(\ref{transfo}). With the latter result (\ref{transfo}), 
we can take any gauge and thus
will consider that the external potential in linear. We note that
in this case the phase can be written

\be\label{remark}
g\int_x^y dz^\mu A^{ext}_\mu(z)=\frac{g}{2}(y^\mu-x^\mu)A^{ext}_\mu(x+y)
\ee

\nin We will check now that the same phase dependence for the
complete fermion propagator and vertex is consistent with the differential
equation (\ref{evolg}). Let us plug

\bea\label{ansatz}
G(x,y)&=&e^{ig\int_x^y dz^\mu A^{ext}_\mu(z)}\tilde G(x-y)\\
\Lambda^{\mu}(z;x,y)&=&e^{ig\int_x^y dz^\mu A^{ext}_\mu(z)}
\tilde \Lambda^{\mu}(z-x,z-y)\nonumber
\eea

\nin into (\ref{evolg}). We obtain then, after
a change of variable and since the potential is linear,

\bea\label{prems}
&&\left(i\int_x^y dz^\mu A^{ext}_\mu(z)\right)
\tilde G(x-y)+\partial_g\tilde G(x-y)\\
&=&-i\int_{z_1z_2z_3}A_\mu^{ext}(z_1)\tilde G(x-z_2)
\tilde\Lambda^\mu(z_1-z_2,z_1-z_3)\tilde G(z_3-y)\nonu
&=&-i\int_{z_1z_2z_3}A_\mu^{ext}(z_1)\tilde G(-z_2)
\tilde\Lambda^\mu(z_1-z_2-u,z_1-z_3+u)\tilde G(z_3)\nonu
&&~~~~~~~~-iA_\mu^{ext}(v)\int_{z_1z_2z_3}\tilde G(-z_2)
\tilde\Lambda^\mu(z_1-z_2-u,z_1-z_3+u)\tilde G(z_3)\nonumber
\eea

\nin where $u=(x-y)/2$ and $v=(x+y)/2$.
We can write after an integration by parts (omitting the surface term)

\be
\int_{z_1}\tilde\Lambda^\mu(z_1-z_2-u,z_1-z_3+u)=
-\int_{z_1}z_1^\mu\partial^{z_1}_\nu\Lambda^\nu(z_1;z_2+u,z_3-u)
e^{-ig\int_{z_2+u}^{z_3-u} dz^\mu A^{ext}_\mu(z)}
\ee

\nin which becomes, with the Ward identity (\ref{ward}),

\be\label{simpl}
\int_{z_1}\tilde\Lambda^\mu(z_1-z_2-u,z_1-z_3+u)=
(z_3^\mu-z_2^\mu-2u^\mu)G^{-1}(z_2+u,z_3-u)
e^{-ig\int_{z_2+u}^{z_3-u} dz^\mu A^{ext}_\mu(z)}
\ee

\nin Taking into account (\ref{simpl}), we simply obtain for the second
integral of the right-hand side of (\ref{prems})

\be
\int_{z_1z_2z_3}\tilde G(-z_2)
\tilde\Lambda^\mu(z_1-z_2-u,z_1-z_3+u)\tilde G(z_3)=2u^\mu G(u,-u)
=(x^\mu-y^\mu)\tilde G(x-y)
\ee

\nin where we used the fact that $\int_{-u}^u dz^\mu A^{ext}_\mu(z)=0$
since $A^{ext}$ is linear. We see now with the help of (\ref{remark})
that the terms which do not depend only on
the difference $x-y$ cancel in (\ref{prems}), leading to the
consistent $(x-y)$-dependent differential equation (we note $z=x-y$)

\be
\partial_g\tilde G(z)+i\int_{z_1z_2z_3}A^{ext}_\mu(z_1)\tilde G(-z_2)
\tilde \Lambda^\mu\left(z_1-z_2-\frac{z}{2},z_1-z_3+\frac{z}{2}\right)
\tilde G(z_3)=0
\ee

\nin
Thus the ansatz (\ref{ansatz}) verifies the differential
equation (\ref{evolg}).
If we take another gauge which is not linear,
the relation (\ref{transfo}) will lead
us to the same conclusion (\ref{ansatz}).

To conclude, we stress that the linear functional differential equation
(\ref{evoll}) leads to general considerations concerning the gauge
dependence. This equation expresses the
invariance of the functional form of the effective action with
respect to a change in the external field amplitude. 
Its differentiation with respect to the
fields leads to relations that have to be satisfied
between $n-$point and $(n+1)-$point
proper graphs and in this respect, these relations play a role
similar to Ward identities.
More specific informations
concerning the effects of an external field on the fermion or
photon dynamics (dynamical mass generation, photon splitting,
electron-positron pair creation,...) should be obtained with the
help of the equation (\ref{evol}) and this work is under study.

\section*{Acknowledgements}

This work has been done within the TMR project 'Finite temperature
phase transitions in particle Physics', EU contract number:
FMRX-CT97-0122. I would like to thank J.Polonyi for introducing me
to the functional method described here. I would like to thank
also K.Farakos, G.Koutsoumbas, P.Pasipoularides and A.Kehagias for
useful discussions, as well as C.Schubert and H.Gies for
correspondence.

\end{document}